\documentclass[aps,superscriptaddress, preprint, longbibliography]{revtex4-1}

\usepackage{hyperref}
\usepackage{latexsym}
\usepackage{amsmath}
\usepackage{amsthm}
\usepackage{amssymb}
\usepackage{amsfonts}
\usepackage{bm}
\usepackage{revsymb}

\usepackage{graphicx}
\usepackage[caption=false]{subfig}

\usepackage{graphicx}

\newcommand{\be}{\begin{equation}}
\newcommand{\ee}{\end{equation}}
\newcommand{\ba}{\begin{array}}
\newcommand{\ea}{\end{array}}
\newcommand{\bea}{\begin{eqnarray}}
\newcommand{\eea}{\end{eqnarray}}

\newcommand{\ket}[1]{| #1 \rangle}
\newcommand{\bra}[1]{\langle #1 |}

\begin{document}


\title{ Adiabatic and Hamiltonian computing on a 2D lattice with simple 2-qubit interactions}

\author{Seth Lloyd}
\affiliation{Department of Mechanical Engineering,
Massachusetts Institute of Technology, Cambridge, USA}
\author{Barbara M. \surname{Terhal}}
\affiliation{JARA Institute for Quantum Information, RWTH Aachen University, 52056 Aachen, Germany}

\date{\today}

\begin{abstract}
We show how to perform universal Hamiltonian and adiabatic computing using a 
time-independent Hamiltonian on a 2D grid describing a system
of hopping particles which string together and interact 
to perform the computation.  In this construction, 
the movement of one particle is controlled by 
the presence or absence of other particles, an effective quantum
field effect transistor that allows the construction
of controlled-NOT and controlled-rotation gates.  
The construction translates into a model for universal quantum computation with 
time-independent 2-qubit ZZ and XX+YY interactions on an (almost) planar 
grid. The effective Hamiltonian is arrived at by a single use of first-order 
perturbation theory avoiding the use of perturbation gadgets. The dynamics 
and spectral properties of the effective Hamiltonian can be fully determined 
as it corresponds to a particular realization of a mapping between a 
quantum circuit and a Hamiltonian called the space-time 
circuit-to-Hamiltonian construction.  Because of the simple interactions
required, and because no higher-order perturbation gadgets are employed,
our construction is potentially realizable using superconducting or
other solid-state qubits. 
\end{abstract}

\maketitle

\tableofcontents

\section{Introduction}

The first proposals for quantum computers used time-dependent Hamiltonians
to enact unitary quantum logic gates \cite{Benioff1980, Benioff1982,
Deutsch1985, Lloyd1993, Monroe1995} and the first prototype
quantum computers were realized using such time-dependent methods via
electromagnetic resonance \cite{Cory1997, Chuang1998}.  
In 1986, Feynman proposed a 
method for performing quantum computation using a time-independent 
Hamiltonian \cite{feynman:qmc}.
His motivation was to make a model of quantum computation that resembled
more closely the time-independent Hamiltonian dynamics of the fundamental
laws of physics.  Feynman's trick was to adjoin a global clock variable
that regulated the pace of the computation.   
In 1987, Margolus constructed a model for Hamiltonian
quantum computation that was spatially homogeneous,
eliminating the pointer variable by embedding the computation in an
asynchronous cellular automaton \cite{margolus}.  In the Margolus method,
the role of the pointer variable is subsumed in the positions 
of clock particles or `chronons' that carry quantum bits with 
them as they progress through the computation. 
The global clock variable is thus replaced by
a local clock variable, one for each degree of freedom in the computation.

The Feynman and Margolus models were
originally conceived as dynamic models in which the clock
or the chronons are prepared in a traveling wave state that propagates
through the computation. Later analyses of these models have shown that it is not necessary to prepare
the initial state in traveling wave state: one can just initialize the input state at fixed
initial clock time (see e.g. \cite{nagaj:circular}).

It was also noted that the Feynman Hamiltonian allows quantum computation
to be embedded in the ground state of the quantum system; it provides
a means for mapping the time-dynamics of a quantum system or a quantum circuit  
onto the ground-state of a master `circuit' Hamiltonian which includes the dynamics of the clock variable \cite{KSV:computation}.
This ground state can then be reached via adiabatic quantum computation \cite{ADLLKR:adia}.  The practical disadvantage of the universal circuit Hamiltonian obtained through this construction
is that it involves many-qubit interactions or qudit degrees of freedom. Such interactions can rewritten in terms of simpler, say,
2-qubit interactions through the use of perturbation theory (by using so-called perturbation gadgets \cite{kkr:hamsiam, OT:qma, BDLT:pert_gen, BL:adia, Babbush}). However, effective interactions obtained in $k$th-order degenerate perturbation theory with perturbative coupling $g$ and gap $\Delta$ of the unperturbed Hamiltonian scale in strength as $g (g/\Delta)^{k-1}$ leading to a correspondingly small gap of the effective Hamiltonian (as compared to the physical device temperature). In addition, multiple uses of (higher-order) perturbation theory lead to Hamiltonians with undesirable qubit overhead and complexity. 
Consequently, existing models of Hamiltonian quantum computation 
based on pairwise qubit interactions are not particularly suitable
for physical implementation using, e.g., solid-state quantum
information processors. 


The Margolus asynchronous cellular automaton model of Hamiltonian
quantum computation relies on spatially homogeneous interactions
which allow the chronons that carry the computation 
to progress at different rates at different points: this construction
can be thought of as a quantum-computation based model of 
Wheeler's `many-fingered time' \cite{MisnerThorneWheeler},
and has been recently formalized in \cite{BT:spacetime} 
under the name space-time circuit-to-Hamiltonian construction. 
Several works have either explicitly or implicitly formulated proposals 
of doing universal Hamiltonian computation \cite{janzing:pra, CGW:walk} or quantum adiabatic computation \cite{MMC:groundstate, mizel:ft,MLM:qadiabatic, GTV:adia} using this construction. A related proposal which seeks to do adiabatic computation using the idea of quantum adiabatic transistors has been formulated in \cite{BFC:trans, BF:adia_tele}. Ref.~\cite{nagaj} has proposed a way of doing Hamiltonian computing using the Feynman construction using only 2-qubit interaction and no application of perturbation theory. However, in order to make this model geometrically local on a 2D grid, the author estimates that each qubit is involved in a high number, at most 28, qubit interactions. In addition, the use of the global Feynman clock leads to an inefficient usage of space (i.e., number of qubits) and time resources in Hamiltonian computing:  the gates in the original circuit are executed sequentially in the Hamiltonian computation so that the time-duration of the computation scales as a polynomial in the size of the original circuit.  In contrast, in the space-time construction the duration of the Hamiltonian or adiabatic computation scales with the depth of the original quantum circuit that the computation simulates. \\

\noindent Our results are an improvement over the previous constructions 
in the following sense: 

The 2D grid Hamiltonian proposed in \cite{janzing:pra} for running an 
autonomous programmable quantum computation is based on a strong attractive 
interaction between hopping spin-$1/2$ particles so that in first-order 
perturbation theory these particles hop together through spatial areas where 
their internal states are changed according to the circuit to be implemented. 
This particular realization uses, similarly to the current work, an 
attractive interaction between particles. If one translates this construction to qubits the author's use of holonomic computation to implement logic results in 4-qubit interactions. If the construction is described with particles with spin, it uses attractive particle interactions, single particle hopping terms and terms which couples the spins of pairs of particles. In addition, our construction goes beyond the results in Refs.~\cite{janzing:pra, nagaj} by showing how to do adiabatic computation using similar 2-qubit interactions and known bounded gap above the ground state.
 
 In \cite{BFC:trans} the resulting Hamiltonian requires particular 4-qubit 
interactions and thus a further use of perturbation theory. Even though we reduce particular 4-qubit interactions to 2-qubit interactions by means of a single use of 1st order perturbation theory, it is unlikely that this trick is possible for arbitrary 4-qubit interactions. The Hamiltonian and adiabatic computation that we propose bears the closest resemblance to the 2D grid model Hamiltonian in \cite{GTV:adia}: we show how mathematical results concerning this construction directly carry over to the results in this paper. The particle interactions in \cite{GTV:adia} (and the original \cite{MLM:qadiabatic}) require pairs of particles to jointly move or hop {\em while updating their internal states}: we expect that a reduction of such interactions to 2-local interactions is inefficient, thus strongly favoring the novel construction in this paper.  Two crucial steps give us this improvement over \cite{GTV:adia}, namely the use of the railroad switch idea \cite{feynman:qmc, nagaj:circular, nagaj} to do a classical CNOT or Toffoli gate, and the use of 1st order perturbation theory to coordinate the joint movement of the particles so that they execute the computation jointly. 

A goal of this paper is to construct models for Hamiltonian and adiabatic quantum computation that are compatible with experimentally available couplings in a physical device such 2D arrays of spin qubits in semi-conducting quantum dots or superconducting qubits.  The recent demonstration of  large-scale quantum annealers using superconducting circuits suggests that adiabatic or Hamiltonian computation might provide a fruitful  method for performing quantum computation.  While existing quantum annealers realize a transverse Ising model system, the more general couplings afforded by superconducting and other solid-state and quasi-solid state systems such as optical lattices might be used to perform general adiabatic quantum computation.  We discuss possible physical realizations in Section \ref{sec:phys}.

Running a solid-state (quantum) computation using a time-independent Hamiltonian has the advantage of requiring no AC fields on chip, thus removing
the challenge of placing active control lines between quantum degrees of freedom and turning interactions on and off. A disadvantage of running the quantum degrees of freedom through stationary, in-place gates is that we trade a time-dependence for a space-dependence and thus a 1D quantum circuit is executed on a 2D grid.  


\subsection{Quantum Field Effect Transistor and Perturbation Theory}

An essential element in our construction is that we use high energy `blocking' terms ($H_0$) to prevent lower energy hopping or kinetic terms ($V$) from operating.  Blocking can be thought of as a coherent quantum version of the field-effect transistor that underlies classical computation: the presence of particles (electrons/holes) in the gate electrode prevents other particles from moving from source to drain.  In classical computation the high-energy terms correspond to a tuning of the classical electrostatic potential by the gate electrode so that a sufficiently high barrier, larger than the kinetic energy of the particle at the source, blocks the forward motion. In the quantum version the potential itself is formed by the presence or absence of a (charged) particle. 
Considering that the hopping particle can always tunnel through the energy barrier, one has to choose the kinetic energy perturbatively weak compared to the height of the barrier.

\begin{figure}[ht]
   \centering
    \includegraphics[width=0.7\hsize]{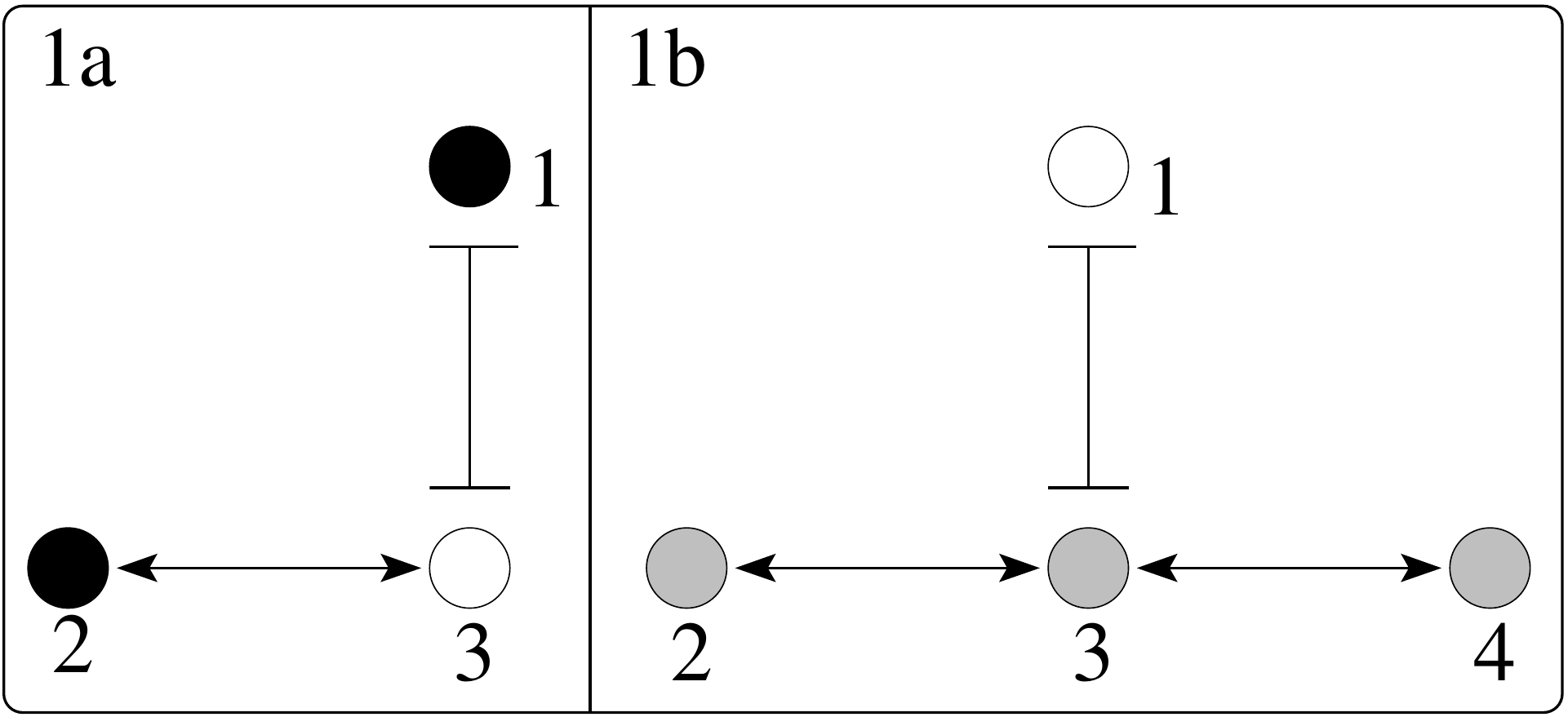} 
       \caption{(a) A pairwise
Hamiltonian induces coherent hopping between sites $2$ and $3$ 
with amplitude $g$.   A pairwise blocking Hamiltonian raises
the energy of a particle at site $3$ to $\Delta$ when there is
also a particle at site 1.  In degenerate perturbation theory,
the sum of these pairwise Hamiltonians results in a low-energy sector with an effective three-local Hamiltonian:
the presence of a particle at site $1$ energetically blocks a second particle
from hopping from $2$ to $3$. (b) Quantum field effect transistor: the
particle at the `source' site $2$ can propagate to the `drain' site $4$
only if the intermediate site $3$ is not blocked.  Tunneling through the
blocked site can be suppressed by adding additional intermediate sites
with high energy.  } \label{fig:1} \end{figure}

We illustrate the idea of blocking with a small example, 
shown in Fig.~\ref{fig:1}(a).   Take the state $|1\rangle$ to represent
the presence of a `particle' on a site (represented in Fig. 1
by a filled circle $\bullet$) and the state $|0\rangle$ to represent
the absence of a particle (represented by an empty circle $\circ$).
Consider a Hamitonian $H_{123} =  H_0+g V$ where $H_0=\Delta \ket{11}\bra{11}_{13}$ and $V= - (\sigma_3^+ \sigma_2^- +\sigma_3^- \sigma_2^+)$.
If $\Delta\gg g$ the term $H_0$ `blocks' a particle from making the 
transition from site 2 to site 3 if another particle already occupies site 1.
$H_{123}$ has a simple block-diagonal form in the basis $\ket{0}\bra{0}_1$ and $\ket{1}\bra{1}_1$, that is
\begin{equation}
H_{123}= \left(\begin{array}{cc} g V & 0 \\ 0 & \Delta \ket{1}\bra{1}_3  +g V\end{array}\right), 
\end{equation}
In the $\ket{0}\bra{0}_1$ part of the low-energy sector the Hamiltonian is simply $g V$. In the $\ket{1}\bra{1}_1$ sector the eigenstates on sites 2 and 3 couple the unperturbed low-energy state $\ket{10}_{23}$ and high-energy state $\ket{01}_{23}$, i.e $\ket{\psi_{\pm}} \propto \frac{2g}{\Delta}\ket{01}_{23}+(1\pm \sqrt{1+4 g^2/\Delta^2}) \ket{10}_{23}$ with energy $E_{\pm}=\frac{\Delta}{2}(1\mp \sqrt{1+4g^2/\Delta^2})$. To zero'th order in $g/\Delta$ the low-energy state equals $\ket{\psi_+} \approx \ket{10}_{23}$ while in first-order in $g/\Delta$ $\ket{\psi_+}$ has an amplitude $O(g/\Delta)$ for the high-energy state $\ket{01}_{23}$: the maximum probability of finding the system in the high-energy sector over time is thus suppressed to $O(g^2/\Delta^2)$.   The probability of finding the system in the high-energy sector
could be further decreased by interactions with the environment at temperature $T < E$. We can thus say that to lowest order in $g$, there exists an effective low-energy Hamiltonian whose dynamics is decoupled from the unperturbed high-energy sector and which equals   
\begin{equation}
H^{\it eff}=g \ket{0}\bra{0}_1 \otimes V-\frac{g^2}{\Delta} \ket{1}\bra{1}_1 \otimes \ket{10}\bra{10}_{23}. 
\label{eq:simple_heff}
\end{equation}
We see from this simple example that in the presence of blocking, 2-local hopping terms can take on a conditional nature: a particle can only hop from one site to another if it is not blocked.   As a consequence, the effective low-energy Hamiltonian $H^{\it eff}$ is now 3-local. 
When performed coherently using multiple particles, by the method described below, such blocking allows one to build up a quantum computation out of a sequence of blocking and hopping moves. We note that the idea of creating blocking energy barriers for certain particle occupations is also used in various first-order perturbation gadgets in \cite{BH:pert}. 

In our detailed construction in Section \ref{sec:ham} the hopping of particles is controlled by the presence of particles nearby. In particular, the {\em presence} of a neighbor particle provides a 2-site {\em energy well} for another particle. This means that the energy barrier for a particle to go off by itself is high on all sites where there are no particles nearby and thus tunneling through this long energy barrier is completely suppressed. 


When the blocking site separates a `source' and `drain' site, as in Fig. (1b),
the resulting device makes up a Hamiltonian quantum field effect transistor
(qFET), where the particle at site 2 can move to site 4 only if
site 3 is not blocked.     Just as in a conventional electronic computer,
where field effect transistors are used to guide charged particles
through the computation and to perform logic, the quantum field effect
transistor will be used to guide qubits through the computation and
to perform quantum logic.    The detailed construction of the use
of quantum field effect transistors to perform Hamiltonian quantum 
computation will be given in Section V.   

Note that even when the particle at site 2 is blocked by an energy
barrier at 3, it can still tunnel through this barrier to site 4, 
leading to incorrect dynamics at higher order in perturbation theory.
To suppress such tunneling, we can add a sequence of intermediate
sites over which the energy barrier remains high: the details
of this construction are given in Section IIB. 



For a general analysis of the effect of blocking and weak particle hopping for many particles we can invoke the results of degenerate perturbation theory. Let us briefly summarize the results of the systematic Schrieffer-Wolff peturbative method \cite{BDLT:pert_gen, BDL:pert}. Let $H_0$ be a many-particle Hamiltonian with an energy gap $\Delta$ between a degenerate ground-state sector with, say, zero energy, and higher-energy states. For example, the Hamiltonian $H_0$ will add energy penalties when particles do not stay together, i.e. it is blocking a particle from moving away while other particles stay behind.  Let $P_0$ be the projector onto the degenerate ground state sector of this Hamiltonian.    Let a perturbation $g V$ be added to $H_0$ where $V$ is a sum of hopping terms for the particles and $g \ll \Delta$. One can show \cite{BDL:pert} that there always exist a unitary transformation $U$ such that $U (H_0+V) U^{\dagger}$ is block-diagonal in the original low- and high-energy eigenspaces of $H_0$. One defines a `low-energy Hamiltonian' $\tilde{H}= P_0 U (H_0+V) U^{\dagger} P_0$ which has then, by definition, the same spectrum as part of $H_0+V$. One can define a perturbative expansion for the unitary operator $U$ which gives rise to a Taylor expansion for $\tilde{H}$ in $g/\Delta$. Cutting off this Taylor expansion for $\tilde{H}$ at a given desired, say $k$th, order in $g/\Delta$ then gives us an approximate effective Hamiltonian $H^{\it eff}$. One expects that this approximate effective Hamiltonian gives a proper description of the low-energy dynamics of $U (H_0+V)U^{\dagger}$ up to errors in energy eigenvalues which scale in strength as $g (g/\Delta)^k$ (and scaling extensively with system size). The unitary transformation $U$ can be thought of as a dressing of the unperturbed eigenstates by the perturbation: if we develop the expansion up to $k$th order, these dressed states are correct up to amplitudes of strength $(g/\Delta)^k$.

To implement the blocking ideas of the quantum field effect transistor we only use perturbation theory to lowest order so that
the effective Hamiltonian equals $H^{\it eff}=g P_0 V P_0$, i.e. of strength $g$, and energy corrections are $O(g^2/\Delta)$. At this order $U=I$ and the dressed states are the unperturbed eigenstates of $H_0$. In the simple example of Fig.~\ref{fig:1} this first-order effective Hamiltonian is only the first term in Eq.~(\ref{eq:simple_heff}) while in second-order one obtains the entire expression in Eq.~(\ref{eq:simple_heff}).

In our analysis we will also discuss the effect of higher-order terms in a perturbative expansion of $\tilde{H}$. We can show that such 
higher-order terms do, assuming an otherwise error-free running of the computation, not affect the logic of the quantum computation that is executed, although it will affect the spectral properties of the Hamiltonian, the form of the dressed states and thus the time-dynamics.

It is important to note that errors in the perturbatively derived effective dynamics always scale extensively with system size,  i.e. number of single-particle modes $n$ \cite{BDLT:pert_gen}. This is the reason why in some quantum complexity applications of perturbation theory \cite{OT:qma} it is required that $g ||V|| \ll \Delta$ implying an unphysical scaling of $g \sim 1/n$ as $||V|| \sim n$. Of course, an extensive scaling of errors with computation size, is entirely natural from the point of view of a quantum circuit model in which each gate has a certain error rate so that the total error rate scales with the total number of gates. 
Furthermore, in many physical implementations of quantum computation, two qubit gate interactions are derived using perturbation theory from more basic interactions and even though these qubits all need to couple together and thus perturbation theory should be applied at the level of a many-qubit system, this is never undertaken. The answer to error accumulation in the quantum circuit model is the use of quantum error correction and fault-tolerance so that error rates on logical qubits are suppressed \cite{quantumcodes_review}. We discuss some aspects of making the computation fault-tolerant in Section VI.

\section{Hamiltonian Quantum Computation}
\label{sec:ham}

Here we describe in detail the execution of a one-dimensional quantum 
circuit using a two-dimensional time-independent Hamiltonian with 
nearest-neighbor interactions.   For clarity of exposition, we present
our analysis in terms of hopping of spin-$1/2$ particles.  Each spin-1/2
particle implements a `chronon' that propagates through the computation,
carrying with it an internal qubit on its spin degree of freedom.
In section \ref{sec:dual} we discuss what the terms in the Hamiltonian 
look like when one represents the spin-$1/2$ particles using dual-rail qubits,
so that the state $|00\rangle$ represents the absence of a particle, 
$|10\rangle$ represents the presence of a particle with internal
qubit state $|0\rangle$, and $|01\rangle$ represents the presence
of a particle with internal qubit state $|1\rangle$ (the state
$|11\rangle$ is unused).    In the spin-1/2 hopping model, particles
hop from site to site in the presence of neighboring particles.
As they hop, their internal qubits can change according to $SU(2)$
rotations.   When the path along which hopping takes place
is determined by the internal state of a neighboring particle --
a {\it spin-dependent} quantum field effect transistor -- then the 
internal qubit of the hopping particle can undergo a controlled rotation
or a CNOT controlled by the internal state of the neighbor.
Although the goal of this paper is to show how Hamiltonian quantum
computation can be implemented using pairwise interactions between
qubits, the construction given in this section using pairwise
interactions between spin-1/2 particles also represents a viable
path to implementable Hamiltonian quantum computation, using,
e.g., spin-dependent electron tunneling \cite{vanderseypen04}. 

\begin{figure}[htb]
 \includegraphics[width=0.4\hsize]{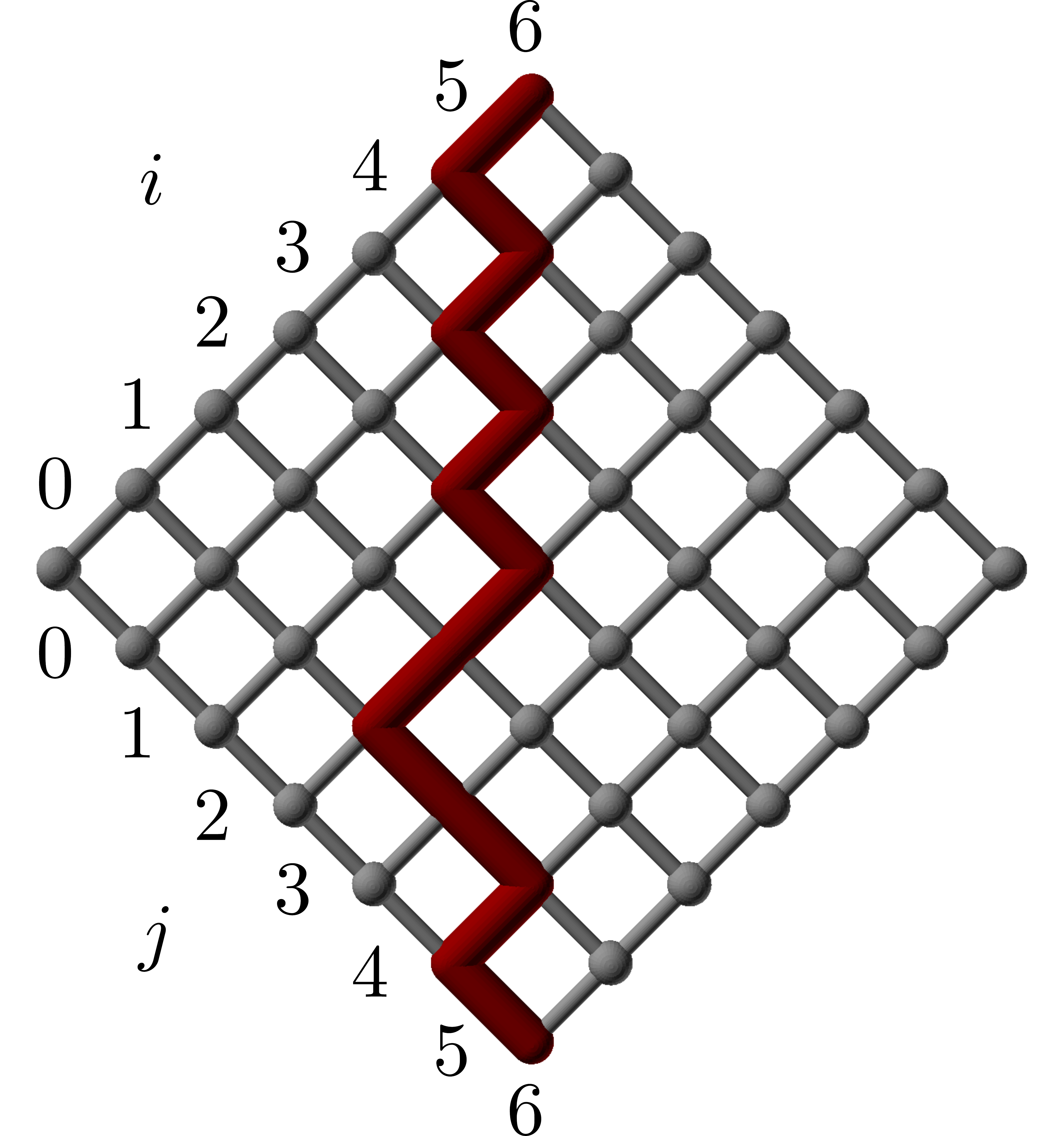}
\caption{\label{fig:rotated_grid}Sites are labeled $(i,j)$ with $i,j\in \{0,\ldots,m\}$. The total number of particles is $2m+1$ counting the stationary dummy particles at the top and bottom of the grid.  There is a single particle on each horizontal line of sites. The red string denotes the locations of the particles and represents a partially-completed computation. The string can be described by a bit string $z$ of length $2m$, with Hamming weight $m$, when we identify $/=0$ and $\backslash=1$.}
\end{figure}

For concreteness, we consider a rotated 2D grid depicted in  Fig.~\ref{fig:rotated_grid}. At each site $(i,j)$, $i,j\in \{0,\ldots,m\}$ on this lattice a particle with spin $1/2$ can reside. There is assumed to be only one particle per horizontal line so in total $2m+1$ particles: the dynamics of the Hamiltonian preserves this property and will let particles move along the horizontal lines, hopping from site to site. At the top and bottom of the lattice, i.e. sites $(0,m)$ and $(m,0)$, the particle is therefore fixed in position and undergoes no dynamics.

Initially, at the beginning of the computation, the particles are all residing on the left end of the grid, one on each site. Using nearest-neighbor interactions on this grid, one can embed a one-dimensional quantum circuit with nearest-neighbor interactions (see \cite{janzing:pra, GTV:adia}). Each horizontal line then represents a single qubit wire of the one-dimensional quantum circuit. Using SWAP gates such circuit can be used to run an arbitrary quantum computation. Instead of using this rectangular grid one can imagine a line of particles hopping forward over a grid, ---a spatial execution of a one-dimensional quantum circuit--. It was noted however in \cite{janzing:pra, GTV:adia} that if the embedded quantum circuit is relatively small compared to $m$ so that it can be embedded in the {\em expanding} region of the grid where particles are gradually added at the boundaries, then the forward motion of the particles in the Hamiltonian computation is very efficient. In essence the boundary condition imposed by this grid breaks the time-reversal of the computation: the string is more likely to move from the boundary to the bulk as the number of bulk strings is much larger. Furthermore, both adiabatic and Hamiltonian computation are easy to analyze in this geometry.

The entire Hamiltonian of the system will be $H=H_{\rm string}+g V_{\rm hop}$ where $H_{\rm string}$ is a strong term which enforces particles to stay close together and form a connected string. In other words, $H_{\rm string}$ gives a penalty to particle configurations which do not form a connected string over the lattice as in Fig.~\ref{fig:rotated_grid}. $H_{\rm string}$ will have a degenerate ground space with zero energy separated by a gap $\Delta$ to higher excited `broken string' states.  In $H_{\rm string}$ each particle interacts via a strong Ising-like interaction with four other particles, namely its neighbor particles on the rotated grid. The perturbative parameter $g \ll \Delta$ and $V_{\rm hop}$ is a sum of hopping (kinetic) terms which move particles forward or backwards from site to site (on a horizontal line). 

If we treat $V_{\rm hop}$ perturbatively, the effective Hamiltonian in $1$st order perturbation theory equals 
$H^{\it eff}=P_{\rm string} V_{\rm hop} P_{\rm string}+O(g^2/\Delta)$. This effective Hamiltonian is thus comprised of hopping terms which preserve the connectedness of the string. Higher-order terms in a perturbative expansion correspond to multiple hops of one or more particles which, taken together, preserve the connectedness of the string. \\

\begin{figure}[htb]
    \centering
    \includegraphics[width=0.6\hsize]{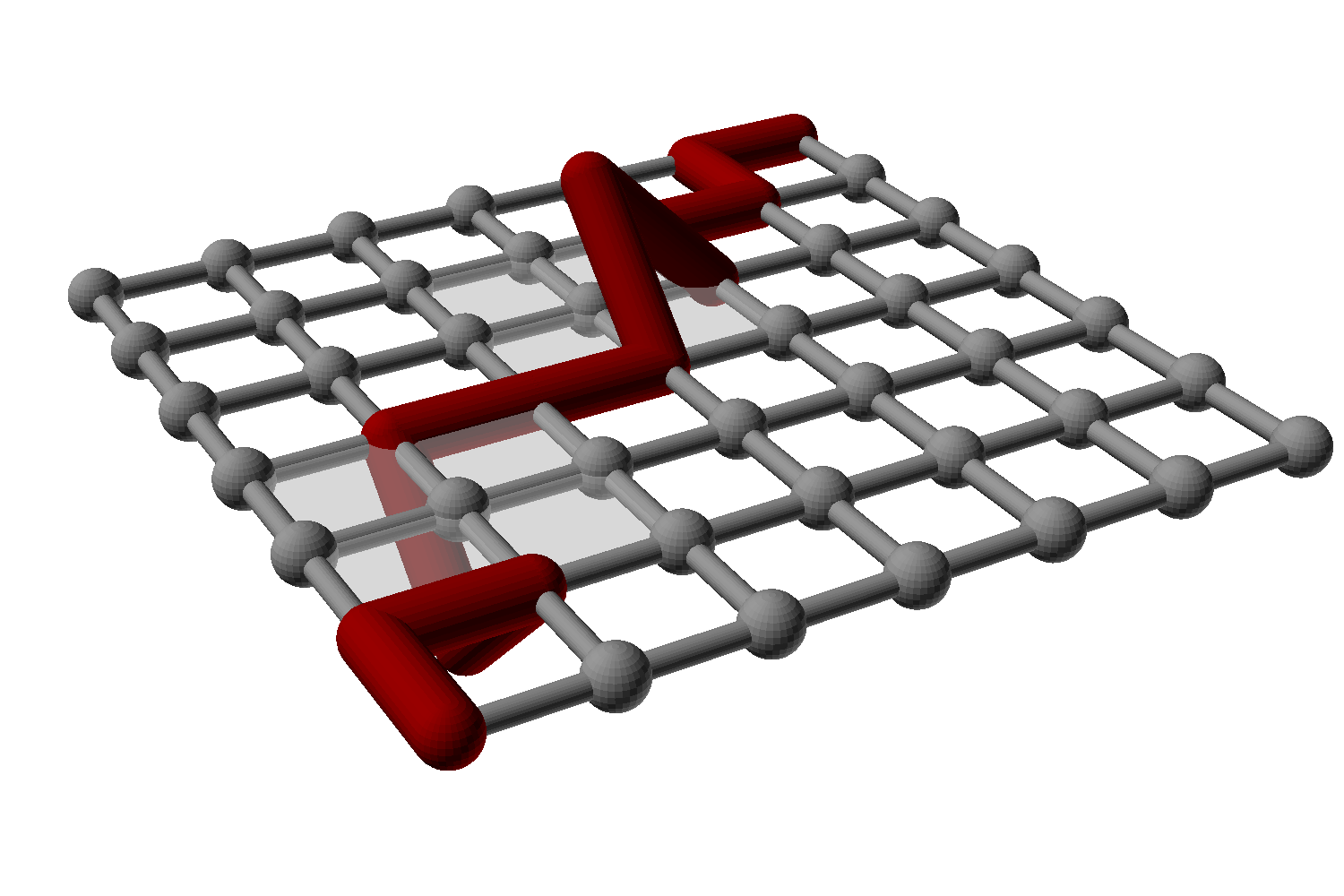}
         \caption{The red string denoting the current state of the computation runs from top to bottom of the lattice through a set of sites. Two light grey CNOT regions, each comprised of four plaquettes,< are depicted. At the center site of these regions the string can either go below or above the plane: which path the string is allowed to take depends on the state of the control qubit of the CNOT gate.}
\label{fig:CNOT3D}
\end{figure}

We will first show how one can execute a simple circuit comprised of single-qubit gates with $H$. In order to include gates such as a CNOT (or controlled-$U$ gate), some of the sites on the grid will be replaced by pairs of sites, for example, one sitting below the 2D plane and another one above. At such locations, the string has the possibility of {\em splitting} and running through either of the sites, below or above the plane, see an example in Fig.~\ref{fig:CNOT3D}.  We will modify $H_{\rm string}$ so that depending on the internal (spin) state of a control particle, only one of these choices has zero energy. Thus $H_{\rm string}$ has zero energy for all connected {\em correct} strings where the correctness depends on the state of some control particles. Similarly, $V_{\rm hop}$ is modified to allow particles to hop to the newly defined sites above and below the plane and undergo internal dynamics which is different for when they hop to a site below or above the plane, thereby allowing controlled-rotations and CNOTs. The string is thus being routed in two different ways capturing the idea of a railroad switch which temporarily routes a train along two different paths. During the quantum computation while the control particles are in superpositions of different internal states, one thus works with a coherent superposition of strings routed partially above or below the plane.

The standard procedure to analyze the dynamics of the (space-time) circuit-to-Hamiltonian construction is to define a unitary transformation which transforms away the internal dynamics of the spin degrees of freedom leaving only the dynamics for the pointer or multiple-pointer/string degrees of freedom, see e.g. \cite{BT:spacetime}. The difference with the previous constructions is the coupling between the internal spin degree of freedom and position of the string used in the CNOT or Toffoli gate. In Appendix \ref{sec:rot} we show how to rotate away the internal dynamics, both for Hamiltonian computation as well as for the quantum adiabatic model introduced below. It then follows that the string dynamics due to $H^{\it eff}$ is easy to represent as one can parametrize a string in Fig.~\ref{fig:rotated_grid} by a bit-string $z$ of length $2m$ and Hamming weight $m$, where bit $0=/$ (resp. $1=\backslash$). For example, the initial string on the left $\ket{z_{\rm init}}$ is the configuration $\ket{00\ldots 0 1\ldots 1}$.  The dynamics of the string in the Hamiltonian computation (resp. adiabatic computation), and therefore the forward motion of the computation, can then be unitarily mapped onto a one-dimensional XY model (resp. Heisenberg model) and this dynamics has been considered previously in \cite{GTV:adia, janzing:pra}. Given that one initializes the computation with a string on the left, the probability of finding the string elsewhere on the lattice after a time-evolution $\exp(-i H^{\it eff} t)$ is also known \cite{GTV:adia}. 

The Hamiltonian that we present assumes that there exists only a single-particle or chronon on each line. It is not possible to enforce this condition locally, but we will show at the end of the paper, Section \ref{sec:loopy}, that one can add simple local interaction terms (which translate as two-qubit ZZ terms in the dual-rail encoding) which ensure that states with multiple particles on a line are of higher energy.


 \subsection{Single-Qubit Gates}
  \label{sec:single}
  It is useful to define the particle number operator at site $(i,j)$ as ${\bf n}[i,j]=\sum_{s=0,1} n_s[i,j]$ where $n_s[i,j]$ is the number operator for the particle at site $(i,j)$ in internal state $s=0,1$ (e.g. spin $\uparrow$ and $\downarrow$), that is, $n_s[i,j]=a^{\dagger}_s[i,j] a_s[i,j]$ where $a^{\dagger}_s[i,j]$ ($a_s[i,j]$) is the creation (annihilation) operator for a particle in internal state $s$ at site $(i,j)$. Sometimes we also write ${\bf n}[v]$ or $a[v]$ etc. where the $v$ label just stands for site $v$, to avoid cumbersome notation.
  We define 
  \begin{equation}
  H_{\rm string}=\frac{\Delta}{4} \sum_e H_e + H_{\rm boundary},
\end{equation}
  where $e$ labels all the edges $e$ of the 2D grid in Fig.~\ref{fig:rotated_grid}. $H_{\rm boundary}$ is a boundary term which acts only on the particles at the boundary of the lattice. The goal is for $H_{\rm string}$ to have a degenerate ground space where each state corresponds to a connected string of particles over the lattice as in Fig.~\ref{fig:rotated_grid}. It is important to remember that we always work in the subspace where there is a single particle per horizontal line. We take an Ising-like interaction per edge $e$:
  \begin{equation}
  H_e=-(I-2 {\bf n}[v])(I-2 {\bf n}[v'])+I.
  \label{eq:edge}
  \end{equation}
 A state with 2 particles at $v$ and $v'$ (${\bf n}[v]={\bf n}[v']=1$) or a state with no particles at $v$ and $v'$ (${\bf n}[v]={\bf n}[v']=0$)
  have energy equal to zero with respect to $\frac{\Delta}{4} H_e$ and a configuration with ${\bf n}[v] \neq {\bf n}[v']$ has energy $\frac{\Delta}{2}$. A particle at a site $v$ in the bulk of the lattice participates in four edge terms $H_e$. For a connected string going through this site, two out of the four terms will have zero energy while the other two edges together give an energy penalty $\Delta$. However at boundary sites, which connect to only 3 other sites, this penalty for a connected string becomes $\frac{\Delta}{2}$ and for the 4 corner sites of the grid a connected string going through these sites picks up no penalty at that site. Thus in order for all connected strings to have equal energy with respect to $H_{\rm string}$, we need to add $\frac{ \Delta}{2}$ penalties at the boundary sites. One takes $H_{\rm boundary}$ as 
  \begin{equation}
  H_{\rm boundary}=\frac{\Delta}{2} \left(\sum_{v \in \mbox{\small boundary}} {\bf n}[v]+\sum_{v \in \mbox{\small corners}} {\bf n}[v]\right).
  \label{eq:bound}
  \end{equation}
 A connected string will then have energy $(2m+1)\Delta$. For a broken string at least two sites will be end-points of a string: such string thus has energy $\Delta$ above the ground space of connected strings. This establishes that the gap of $H_{\rm strings}$ is $\Delta$ and the degenerate ground space consists of connected strings (independent of the spin-state of the particles) with projector $P_{\rm string}$ onto this ground space. 
 
 To implement some single-qubit gates $U$ one takes $V_{\rm hop}=\sum_p V_{\rm hop}^p$ where $p$ runs over plaquettes $p$ of the grid and each plaquette corresponds to some single qubit gate $U$ or $I$. The term $V_{\rm hop}^p$ for a plaquette with corner sites $(i,j),(i+1,j),(i+1,j+1),(i,j+1)$ lets a particle at site $(i,j)$ hop to site $(i+1, j+1)$ while it is changing its internal spin state according to the gate $U$, i.e.
 \begin{equation}
 V_{\rm hop}^p=- \sum_{s,s'=0,1} \bra{s'} U \ket{s} a_{s'}^{\dagger}[i+1,j+1] a_s[i,j]+ {\rm herm. conj.}.
 \label{eq:single}
 \end{equation}
 This form of executing single-qubit gates has been first proposed in \cite{MMC:groundstate}. When $U=I$ gate, this interaction is a simple kinetic energy term of moving particles/fermions on a 1D line. 
  If we treat $H_{\rm string}+g V_{\rm hop}$ perturbatively,
 the effective Hamiltonian reads 
 \begin{eqnarray}
 H^{\it eff}=g P_{\rm string} V_{\rm hop} P_{\rm string}=g \sum_{p} H_{\rm cond. hop}^p+O\left(\frac{g^2}{\Delta}\right), \nonumber \\ 
 H_{\rm cond.hop}^p=- {\bf n}[i+1,j] {\bf n}[i,j+1] \left(\sum_{s,s'=0,1} \bra{s'} U \ket{s} a_{s'}^{\dagger}[i+1,j+1] a_s[i,j]+ {\rm herm. conj.}\right).
 \label{eq:effham1}
 \end{eqnarray}
 In words: particles can only hop forward or backward over a plaquette when there are particles at the top and bottom of the plaquette so that hopping keeps the string connected.  It has been shown in \cite{janzing:pra, GTV:adia} that the string dynamics induced by such effective Hamiltonian $H^{\it eff}$ can be unitarily related to an XY model.
  
In principle, it is possible to give an effective Hamiltonian in the string subspace which includes higher-order virtual hopping processes from string subspace to high-energy disconnected string subspace and then back to string subspace. Such effective Hamiltonian in the low-energy string subspace realizes the proper logical single-qubit gates, but the string dynamics will not be described by an XY model. For example, in second-order in $g$, one has the following possible processes: (1) twice application of some $V_{\rm hop}^p$ leading to no string motion, or a term proportional to $I$, (2) the application of two different terms $V_{\rm hop}^p$ and $V_{\rm hop}^q$ with $q\neq p$, each of which keeps the string connected leading to terms which are unitarily equivalent to $H_{XY}^2$, and (3) the application of two adjacent terms $V_{\rm hop}^p$ and $V_{\rm hop}^q$ with $q\neq p$, such that the application of either one breaks the string but together they map back onto a connected string. This third process has a prefactor $g^2/\Delta$ and does not lead to terms which are equivalent to $H_{XY}^2$.


\begin{figure}[htb]
    \centering
  \includegraphics[width=0.4\hsize]{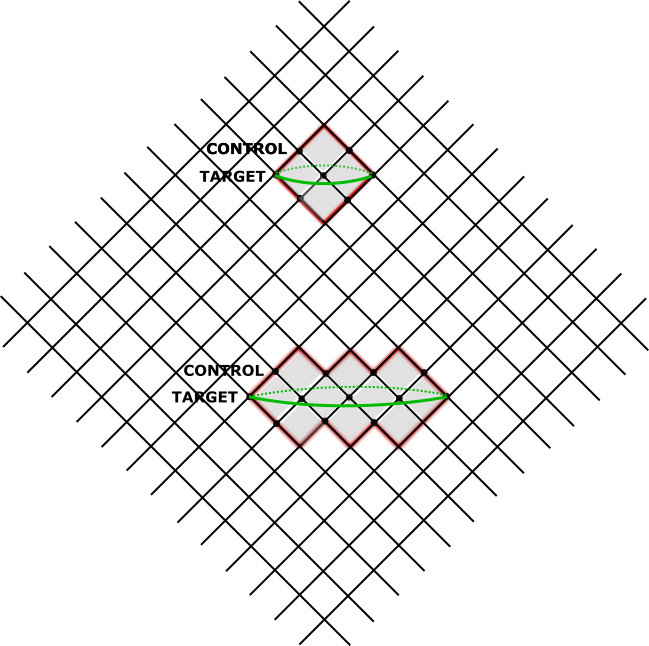}
       \includegraphics[width=0.5\hsize]{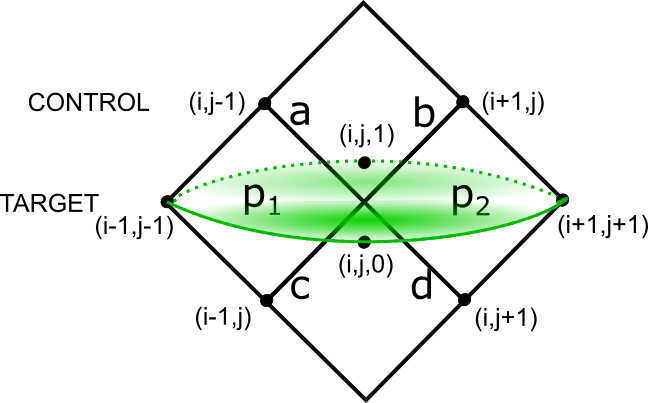}
        \caption{Left: Grey CNOT regions in the rotated grid. In a CNOT region terms in the Hamiltonian are modified in order to execute a CNOT. A long CNOT region is shown at the bottom. Right: detailed picture of the basic CNOT region. In the middle of the CNOT region the state space is doubled (the target particle can hop to a site below or above the plane). Depending on the internal state of the control particle at site $(i,j-1)$ and $(i+1,j)$ an energy penalty is assigned when the target particle is present on the wrong side of the plane. Edges which involve these two central sites at $(i,j,0)$ and $(i,j,1)$ are labeled $a,b,c$ and $d$. The horizontal hopping terms labeled by plaquette $p_1$ and $p_2$ towards the doubled $(i,j,0/1)$ site are modified to execute the proper logic on the target qubit.}
\label{fig:CNOT}
\end{figure}

 \subsection{CNOT Gate}
   \label{sec:CNOT}
We describe the modification of the Hamiltonians $H_{\rm string}$ and $g V_{\rm hop}$ which one allow one to perform controlled-$U$ gates. We illustrate the construction with a CNOT gate. In order to execute a CNOT we will modify a region of four adjacent plaquettes on the lattice; we call this region a CNOT region, see Fig.~\ref{fig:CNOT}. The central site $(i,j)$ in this region will have a doubled state space; there is a site (say) above and below the plane. The (annihilation) operators for this site thus have one additional new label, i.e. they are $a_s[i,j,k=0,1]$ where the $k$ label is the new coordinate. In order to give the modified Hamiltonian, we can use number operators $n_s[i,j,k]=a_s^{\dagger}[i,j,k] a_s[i,j,k]$, ${\bf n}[i,j,k]=\sum_s n_s[i,j,k]$, ${\bf n}[i,j]=\sum_{k=0,1} {\bf n}[i,j,k]$ etc. We can imagine placing such a CNOT region in various places on the grid. 

We modify the edges which are internal to the CNOT region in $H_{\rm string}$: these are the edges $a,b,c$ and $d$ in Fig.~\ref{fig:CNOT}(b). We choose the control qubit above the target qubit. The control qubit is just undergoing the $I$ gate when hopping from $(i,j-1)$  to $(i+1,j)$ in Fig.~\ref{fig:CNOT}. The qubit on the horizontal line below the target qubit and the qubit on the line above the target qubit are not participating in the CNOT gate and could be undergoing single-qubit dynamics or participate in another CNOT gate. Thus the edges $c, d$ are the same as in Eq.~(\ref{eq:edge}), e.g. $H_{c}=-(I-2 {\bf n}[i-1,j])(I-2 {\bf n}[i,j])+I$ where ${\bf n}[i,j]$ is total number operator of the doubled site $(i,j)$ defined above. 

Instead of Eq.~(\ref{eq:edge}), one has for edge $a$ and $b$
\begin{eqnarray}
H_{a} & =&  -(I-2 n_{s=0}[i,j-1])(I-2 {\bf n}[i,j,k=0]) \nonumber \\
& &- (I-2 n_{s=1}[i,j-1])(I-2 {\bf n}[i,j,k=1])+ 2 I, \nonumber \\
H_{b}& = & -(I-2 n_{s=0}[i+1,j])(I-2 {\bf n}[i,j,k=0]) \nonumber \\
&& - (I-2 n_{s=1}[i+1,j])(I-2 {\bf n}[i,j,k=1])+ 2 I.
\label{eq:mod_edge}
\end{eqnarray}
What is the spectrum of, say, $H_a$? Remember we assume that there is at most one particle per horizontal line. When there are no particles on sites $(i,j-1)$ and $(i,j,k)$ for $k=0,1$ $H_a$ has zero energy. When there is a particle at $(i,j-1)$ but no particle at $(i,j,k=0)$ or $(i,j,k=1)$, then $\frac{\Delta}{4} H_a$ equals $\Delta/2$. When ${\bf n}[i,j,k=1]=1, n_{s=1}[i,j-1]=1$, then $ n_{s=0}[i,j-1]={\bf n}[i,j,k=0]=0$, hence the energy is again zero. When ${\bf n}[i,j,k=1]=0$ but $n_{s=1}[i,j-1]=1$, we have $n_{s=0}[i,j-1]=0$ and two choices namely ${\bf n}[i,j,k=0]=0$ or ${\bf n}[i,j,k=0]=1$. In the first case, the energy is zero, the string simply does not yet run over the edge $a$. In the second case, the energy of $\frac{\Delta}{4} H_a$ is $\frac{\Delta}{2}$, and we can say that the string runs {\em incorrectly} over edge $a$: even though the control qubit is in state $s=1$, the target particle is found at site $(i,j,k=0)$. These arguments imply that not continuing along edge $a$ has the same energy penalty as incorrectly continuing along edge $a$. The same arguments apply to the modified edge $b$.
This means that the ground space of $H_{\rm string}$ is formed by all correct strings which have energy $(2m+1)\Delta$ as before. We  denote the projector onto the ground space of $H_{\rm string}$ as $P_{\rm correct\; string}$.  What is the energy of broken or incorrect strings and thus the gap of the modified string Hamiltonian? Again a string can be broken such that it has $\Delta$ energy above the ground space. A string can also be connected but incorrect on the central site in the CNOT region: in that case the string is picking up an energy penalty only on the incorrect edge which costs $\frac{\Delta}{2}$. The gap of the modified $H_{\rm string}$ is thus $\Delta/2$ (which we can of course rescale to $\Delta$ again).

In order to execute a CNOT gate in a region, the weak hopping terms towards the central now doubled site $(i,j)$ also need to be modified. If the particle passes through the site $(i,j,k=0)$, its internal state should be unchanged as the control particle was in the state $\ket{0}$. If the particle passes through the site $(i,j,k=1)$, its internal state should undergo a bit-flip $X$ (for a controlled-$U$ gate this can be an arbitrary unitary $U$). We thus modify Eq.~(\ref{eq:single}) for plaquettes $p_1$ and $p_2$ as in Fig.~\ref{fig:CNOT}. We take
\begin{eqnarray}
V_{\rm hop}^{p_1} & =  -\sum_{s=0,1} a_{s}^{\dagger}[i,j,0] a_s[i-1,j-1]+ {\rm herm. conj.} & \mbox{\em I gate when routed through 0} \nonumber \\
& 
-\sum_{s=0,1} a_{\overline{s}}^{\dagger}[i,j,1] a_s[i-1,j-1]+ {\rm herm. conj.} & \mbox{\em X gate when routed through 1} \nonumber \\
V_{\rm hop}^{p_2}& = -\sum_{s=0,1} \sum_{k=0,1} a_{s}^{\dagger}[i+1,j+1] a_s[i,j,k]+ {\rm herm. conj.} & 
\label{eq:cnot_hop}
\end{eqnarray}

 With the introduction of the CNOT region(s), the effective Hamiltonian in Eq.~(\ref{eq:effham1}), $H^{\it eff}=g P_{\rm correct\; string} V_{\rm hop} P_{\rm correct\;string}$, is again a sum over controlled-hopping terms $g H_{\rm cond.hop}^p$ with the modification that for all plaquettes $p_1$ and $p_2$ in a CNOT region, the controlled-hopping terms equal 
\begin{eqnarray}
H_{\rm cond.hop}^{p_1} & = & - n_{s=0}[i,j-1] {\bf n}[i-1,j] \left(\sum_{s=0,1}a_{s}^{\dagger}[i,j,0] a_s[i-1,j-1]+ {\rm herm.\;conj.}\right). \nonumber \\
& &  - n_{s=1}[i,j-1] {\bf n}[i-1,j] \left(\sum_{s=0,1}a_{\overline{s}}^{\dagger}[i,j,1] a_s[i-1,j-1]+ {\rm herm.\;conj.}\right). \nonumber \\
H_{\rm cond.hop}^{p_2} & = & - \sum_{k=0,1} n_k[i+1,j]{\bf n}[i,j+1]  \left(\sum_{s=0,1} a_{s}^{\dagger}[i+1,j+1] a_s[i,j,k]+ {\rm herm.\; conj.}\right) 
\label{eq:cond_hop_CNOT}
\end{eqnarray}
These terms represent the execution of the CNOT gate via two steps $p_1$ and $p_2$: the target particle hops from site $(i-1,j-1)$ onto either site $(i,j,k=0,1)$ depending on whether there are particles right above and below it and depending on whether the control particle above it is in the state $0$ or $1$. In the next hopping process $p_2$ the particle does not undergo internal dynamics and it can hop from either site $(i,j,k=0)$ or $(i,j,k=1)$ onto $(i+1,j+1)$, at least when there is a particle in the correct spin state above and a particle in an arbitrary state below it. Of course, the Hamiltonian also includes the reverse conditional hopping process. One can note that these controlled-hopping processes $H_{\rm cond.hop}^{p_1}$ and $H_{\rm cond.hop.}^{p_2}$ will take place sequentially as the particles above and below the target particle have to move forward or backward after one of the process takes place in order to have the next one executed.

We can consider what terms are present in the next order of the perturbative expansion where contributions are coming from twice hopping from the correct string space back to the correct string space. If double hopping involves a single particle, then in order for the string to stay connected (and correct), the particle has to hop backward and forward, inducing no effective dynamics. The only other contributions in second-order in the perturbation $V_{\rm hop}$, are from two different particles hopping so that together they preserve the connectedness of the string. The same is true for all third-order processes as it takes at least four hopping processes in order for the particles to hop/tunnel through a CNOT region.
Thus terms up to $O(g^3/\Delta^2)$ give rise to a change in string dynamics, but the logic is still properly executed. In other words, if the string has traveled through the region where the gates have been applied, then the correct circuit has been implemented. 

A fourth-order process can however lead to incorrect logic in the CNOT gate as follows. Assume the control particle is in the state $\ket{0}$. The target particle hops from site $(i-1,j-1)$ through incorrect site $(i,j,1)$, and then again to $(i+1,j+1)$ and the control particle hops from 
$(i,j-1)$ to $(i+1,j)$, while the particle below hops from $(i-1,j)$ to $(i,j+1)$, so that the string stays connected and is correct after these four processes. In this fourth-order process (of order $O(g^4/\Delta^3)$) the target particle undergoes a bitflip although the control qubit was 0. One can suppress these higher-order virtual tunneling processes by replacing the short CNOT by a {\em long CNOT}. The idea of the long CNOT is to enlarge the CNOT region to an elongated CNOT region in which the central site is a sequence of $L$ sites where the string can only go below or above the plane. The hopping terms of these $L$ sites propagate the particles on one site of the plane or the other.
For such long CNOT with $L$ central sites, it takes $L+1$ hops for the target particle to be on the plane again where the fact that it has taken the wrong route is no longer visible (in other words, it takes $L+1$ hops for the particle to tunnel through the energy barrier). For the string to stay connected, the particles above and below the target particle also have to hop along, over $L$ sites. Hence,
the first term which gives rise to incorrect logic is of order $O\left(g (\frac{g}{\Delta})^{3L+1}\right)$.  

\section{Toffoli Gate and Programmable Circuits}

The principle of the spin-dependent quantum field effect transistor can also be used to do a Toffoli gate or make the quantum circuit programmable.
In the Toffoli gate two particles determine how a third qubit should be routed, above or below plane. One can choose these two particles as a particle above and below the target particle, as in the CNOT region in Fig.~\ref{fig:CNOT}.  For a short Toffoli gate, the modification of $H_{\rm string}$ then involves all four edges $a,b,c,d$, each one of them should depend on the internal state of the control particle(s), as in Eq.~(\ref{eq:mod_edge}). 
 
 In order to make the computation programmable, one can add classical control bits to the grid which can route the particles to regions where they can undergo the desired logic. These bits do not need to participate in the dynamics, i.e. but they can interact with the particles  involved in the computation in the Hamiltonian $H_{\rm string}$ giving penalties for incorrect string configurations.

\section{Quantum Adiabatic Computation}

To execute an adiabatic quantum computation on the grid, we define the Hamiltonian $H(\lambda)$ with $\lambda \in [0,1]$. This construction follows the ideas in \cite{GTV:adia}, but the adiabatic Hamiltonian has to be chosen differently in order for the spectral analysis of \cite{GTV:adia} to apply. Furthermore, for adiabatic computation the previously defined Hamiltonian $H=H_{\rm string}+g V_{\rm hop}$ does not suffice: new interactions have to be included in the adiabatic Hamiltonian and the adiabatic parameter has to be chosen so that the ground state is a (weighted) superposition over all strings, thus encoding the computation.

We will take $H(\lambda)=H_{\rm string}+H_{\rm input}+g H_{\rm circuit}(\lambda)$. Here $H_{\rm input}$ is a term setting the initial internal state of the particles to the correct input state of the circuit: one can simply choose the proper occupation of sites on the left half of the lattice. If these initial states are set correctly, $H_{\rm input}$ has zero energy and thus keeps the degeneracy between all correct strings. We will further neglect $H_{\rm input}$ in the discussion here (see \cite{GTV:adia} for how $H_{\rm input}$ makes the ground state of the quantum adiabatic computation unique). 

We construct  $H_{\rm circuit}(\lambda)$ as a modification of $V_{\rm hop}$. Let us first do this for all plaquettes which correspond to single-qubit gates. Let a plaquette $p$ enable a particle to hop from site $(i,j)$ to $(i+1,j+1)$ while there is a particle at the top of the plaquette, site $(i+1,j)$ and at the bottom of the plaquette $(i,j+1)$. The idea of the quantum adiabatic computation is to gradually turn on the motion/kinetic energy of the string using a parameter $\lambda$. Initially for $\lambda=0$, we want to choose $H_{\rm circuit}(\lambda=0)$ such that the straightest string, with the fewest `wiggles', has the lowest energy. There are two such straightest strings on the left and right boundary of the lattice and we include a term $H_{\rm init}$ in $H_{\rm circuit}(\lambda)$ which favors the initial string on the left. We choose $H_{\rm init}={\bf n}[m,1]+{\bf n}[1,m]$, thus penalizing any string which runs along the top and bottom edges on the right.

In the string subspace, it is clear that a term of the form ${\bf n}[i+1,j] {\bf n}[i,j+1]$ provides an energy penalty when the string runs around the plaquette $p$ of which $(i+1,j)$ and $(i,j+1)$ are the top and bottom sites. More precisely, we will take
\begin{equation}
H_{\rm circuit}(\lambda)=\sum_p \left({\bf n}[i+1,j]{\bf n}[i,j+1]+ \lambda V_{\rm hop}^p \right)+\sqrt{1-\lambda^2} H_{\rm init},
\label{eq:circuitH}
\end{equation}
where $V_{\rm hop}^p$ is as before. The additional terms that we have thus added are diagonal in the string basis. Thus in the string subspace, the effective Hamiltonian just equals
\begin{equation}
H^{\it eff}_{\rm circuit}(\lambda)=g\left( \sum_p \left[{\bf n}[i+1,j]{\bf n}[i,j+1]+ \lambda H_{\rm cond. hop}^p\right]+ \sqrt{1-\lambda^2} H_{\rm init}\right).
\end{equation} 
When the quantum circuit contains CNOT gates, we can define the adiabatic Hamiltonian in the same way, using Eq.~(\ref{eq:circuitH}). For certain plaquettes $p$ in the CNOT region $V_{\rm hop}^p$ is chosen according to Eq.~(\ref{eq:cnot_hop}) and for a doubled-site $(i,j)$ we simply use the definition ${\bf n}[i,j]=\sum_k {\bf n}[i,j,k]$, that is, we use the total particle operator on the site.

It can be shown for this effective Hamiltonian $H^{\it eff}_{\rm circuit}(\lambda)$ that one can rotate away the internal spin dynamics including the dependence on the position on the control qubits, see Appendix \ref{sec:rot}. It then follows that we can represent the action of the effective Hamiltonian  by a XXZ Heisenberg chain as in \cite{GTV:adia} in the string subspace spanned by $2m$-bitstrings $\ket{z}$: 
\begin{equation}
\bra{z} \sum_p  ({\bf n}[i+1,j]{\bf n}[i,j+1]+ \lambda H_{\rm cond. hop}^p) \ket{z}=-\frac{1}{2}\bra{z} \sum_{i=1}^{2m-1} (Z_i Z_{i+1}-I)-\lambda (X_i X_{i+1}+ Y_i Y_{i+1})\ket{z}.
\label{eq:XXZ}
\end{equation}
In addition, the term $H_{\rm init}$ can be represented as a kink boundary condition which allows this model to be exactly solvable: thus all previous results concerning the ground state and the gap in \cite{GTV:adia} apply to this effective Hamiltonian.
        
 \section{Dual Rail Encoding: Required Interactions} 
  \label{sec:dual}

 There are different ways of representing the particle with spin physically: the simplest is a dual-rail encoding using two qubits to represent one spin-$1/2$ particle. We note that the fermionic nature of the particles in the construction is not relevant: each fermionic mode can be represented by a qubit where occupation of the mode is $\ket{1}$ and no occupation of the mode is $\ket{0}$ (see \footnote{This fact follows directly for a construction with only single-qubit gates. Each particle is hopping along a one-dimensional wire and interactions between particles on different wires only use number operators. For a Hamiltonian of such a system the Jordan-Wigner transformation mapping fermions onto qubits (i.e. upon ordering all the fermionic sites $i=1, \ldots, N$ one represents $a^{\dagger}_i \rightarrow Z_1 \ldots Z_{i-1}\sigma_i^+$) does not introduce any long-range $Z$-like interactions. However, the construction contains loops of hopping fermionic terms when CNOT gates are included in the circuit. In a loop, one inevitably has hopping terms $a_i^{\dagger} a_j$ for $j\neq i\pm 1$ which under the Jordan-Wigner transformation become for $i < j $, $a_i^{\dagger} a_j \rightarrow \sigma_i^+ Z_i \ldots Z_{j-1} \sigma_j^-$. However, there is at most one particle in any such loop, which means that if the action of such terms is non-trivial there are no particles at sites $i,\ldots j-1$ and thus $Z_i \ldots Z_{j-1}$ has eigenvalue 1 and can be omitted.} for the argument). This means that the two fermionic modes of a particle with spin can be represented by two qubits. We label these two qubits as qubit 0 and 1.  We can choose to represent a particle in state $s=0$ as $\ket{10}_{01}$, i.e. qubit 0 is $\ket{1}$, particle in state $s=1$ as $\ket{01}_{01}$, no particle present is $\ket{00}_{01}$ and the two-qubit state $\ket{11}_{01}$ is not used as there is at most one particle present.
  
\begin{figure}[htb]
    \centering
    \includegraphics[width=0.5\hsize]{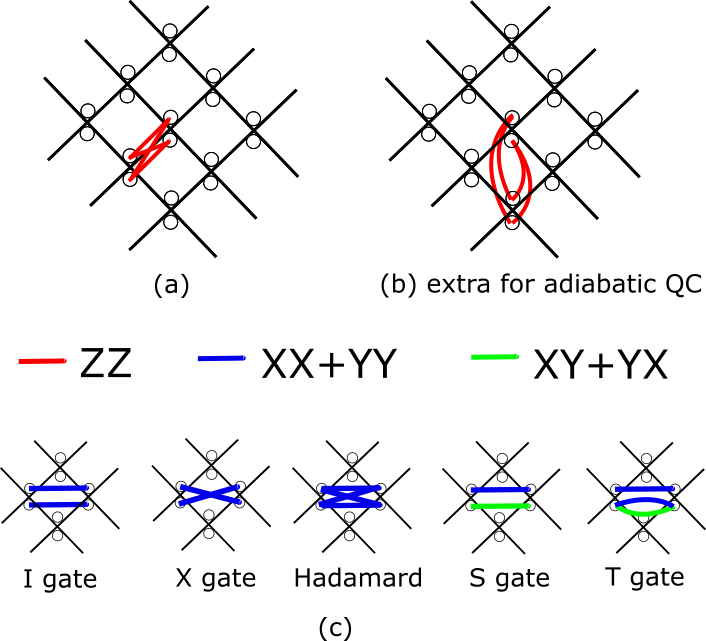}
            \caption{Required two-qubit interactions for Hamiltonian and adiabatic quantum computing. At each site the particle with spin is represented by two qubits ($0$ and $1$) in the dual-rail encoding. (a) For each edge connecting a pair of sites there are 4 strong ZZ interactions (in red) between the four qubits. Thus each qubit interacts with 8 other qubits via ZZ. (b) Only for adiabatic quantum computing there are, in addition, weak ZZ interactions between qubits: for each plaquette the two qubits at the top interact via ZZ with the qubits at the bottom. (c) Single-qubit gates are realized by a plaquette and require weak interactions XX+YY  and/or XY+YX for qubits on the left and right of the plaquette. Here $S={\rm diag}(1,i)$ and $T={\rm diag}(1,\exp(i\pi/4))$. For each plaquette in the total grid which is not used for a non-trivial gate, one has an I gate. The CNOT or Toffoli gate (not shown) use sites above and below the grid, similar strong ZZ edges and an X and I gate.}
\label{fig:connect}
\end{figure}

Thus, each creation operator $a^{\dagger}_s[i,j,k]$ (or annihilation operator $a[i,j,k]$) with spin-label $s$ can be represented as a single-qubit operator $\sigma^+_s[i,j,k]=\ket{1}\bra{0}_{s}[i,j,k]$ (resp. $\sigma^-_s[i,j,k]=\ket{0}\bra{1}_{s}[i,j,k]$). This implies that all hopping terms in $V_{\rm hop}$, Eq.~(\ref{eq:single}), are weak two-qubit terms of the general form
\begin{eqnarray}
\mbox{\em I or X gate } & -(\sigma^+_i \sigma^-_ j+\sigma^-_i \sigma^+_j)=-\frac{1}{2}(X_i X_j+Y_i Y_j) \nonumber \\
\mbox{\em Hadamard gate}&  \frac{\pm 1}{2\sqrt{2}} (X_i X_j+Y_i Y_j) \nonumber \\
\mbox{\em complex single qubit U gate} &  \propto X_i X_j+ Y_i Y_j \mbox{ \em and } \propto X_i Y_j+ X_j Y_i \nonumber \\
\mbox{\em Toffoli or CNOT gate} & -\frac{1}{2}(X_i X_j+Y_i Y_j) 
\end{eqnarray}
In Fig.~\ref{fig:connect}(c) we show these interactions for some specific single-qubit gates.
 
What about the diagonal terms in $H_{\rm string}$ (or $H_{\rm circuit}(\lambda)$, Eq.~(\ref{eq:circuitH}))?

One has $Z_0[v]=I-2 n_{s=0}[v]$ and $Z_1[v]=I-2 n_{s=1}[v]$ for any site $v$. Using this identification, a normal edge $H_e$ in $H_{\rm string}$ as in Eq.~(\ref{eq:edge}) then becomes the Ising interaction 
\begin{equation}
H_e=-(Z_0[v]+Z_1[v]-I)(Z_0[v']+Z_1[v']-I)+I,
\label{eq:dual1}
\end{equation}
hence four pair-wise Ising interactions between the four qubits on sites $v$ and $v'$ (and a sum of local $Z$ terms on all 4 qubits). The boundary terms acting only on qubit sites at the boundary of the lattice, Eq.~(\ref{eq:bound}) can also be represented as single-qubit $Z_0[v]$ and $Z_1[v]$.
A modified edge term in a CNOT region, say, $H_a$ in Eq.~(\ref{eq:mod_edge}), simply equals 
 \begin{equation}
H_a=-Z_0[i,j-1](Z_0[i,j,0]+Z_1[i,j,0]-I)-Z_1[i,j-1](Z_0[i,j,1]+Z_1[i,j,1]-I)+2I.
\label{eq:dual2}
\end{equation}


For adiabatic quantum computing the Hamiltonian in Eq.~(\ref{eq:circuitH}) contains additional interactions which can be represented as weak Z and two-qubit ZZ interactions. In Fig.~\ref{fig:connect} we show the connectivity of the two-qubit interactions for Hamiltonian and adiabatic computing. One can note that qubits connected via Ising ZZ terms are never connected via XX+YY or XY+YX terms and vice versa. For Hamiltonian computing, each qubit $0$ or $1$ at a site $v$ in the bulk of the lattice interacts with 8 other qubits, along the four different edges, via a ZZ interaction. When that qubit is part of the control particle of a CNOT gate, it interacts with 10 other qubits as the state-space is doubled on the central CNOT site.

If we assume that we intersperse single-qubit gates and CNOT gates with I gates, then each qubit interacts in addition with at most $3$ other qubits via XX+YY and XY+YX interactions.  For adiabatic computing each qubit at a site interacts in addition with 4 other qubits (above and below the site), via ZZ, see Fig.~\ref{fig:connect}(b).
 
A possible universal gate set is comprised of the $T={\rm diag}(1,e^{\pi/4})$ gate, a CNOT gate and a Hadamard gate. For the $T$ gate one needs to connect two qubits via a linear combination of XX+YY and XY+YX, see Fig.~\ref{fig:connect}(c), which requires more physical interaction engineering than only XX+YY. Alternatively, one can use Hadamard and the Toffoli gate to get universality \cite{shi:help}. In this case both for the Hamiltonian computation and the adiabatic computation, one only needs the interaction ZZ, Z and $\pm$ XX+YY. The set of interactions ZZ, Z and -(XX+YY) are stoquastic (sign-free) and can be mapped via perturbative reductions onto a transverse field Ising model \cite{BH:pert}. Note that in order to do the Hadamard gate one needs $\pm$ XX+YY and thus the quantum adiabatic Hamiltonian which realizes a universal computation using a Hadamard and Toffoli gate is not stoquastic or sign-free and is unlikely to map onto a transverse field Ising model quantum annealer.

\subsection{Physical Implementations}
\label{sec:phys}

The methods for Hamiltonian quantum computation developed in this paper
are potentially suitable for implementation on multiple physical platforms.
In the form given here, we require (a) strong ferromagnetic ZZ interactions
between neighboring qubits (plus strong local $Z$ terms), and (b) weak XX+YY interactions.    For
example, superconducting transmon qubits may be suitable to 
realize the 2-qubit interactions for Hamiltonian computation 
described in this proposal. The strong ZZ interactions can be obtained by strong capacitively-coupled `octmon' (in analogy with Xmon qubits \cite{barends:xmon}) qubits which have eight arms through which they couple with 
nearest-neighbors octmons. The weak XX+YY interaction between two qubits across a plaquette can be obtained by placing a bus-resonator on the plaquette through which both qubits couple. This resonator-mediated interaction through which the qubits can virtually exchange a photon is the basis of the iSWAP gate \cite{blais:cavity}. The coupling strength $J$ of such term $J (\sigma_i^+ \sigma_j^-+ \sigma_j^- \sigma_i^+)$ scales as $J=g_{JC}^2/\Delta$ where $\Delta$ is the detuning between resonator frequency and qubit frequency (assuming that both qubits are at the same frequency) and $g_{JC}$ is the Jaynes-Cummings coupling. The frequency of each bus-resonator and thus $\Delta$ can be adjusted to fix the strength of the weak coupling $J$. 
  In such a physical set-up one would realize Hamiltonian computation in a rotating frame of each octmon qubit, where the rotating frame is chosen such that one retains single-qubit $Z$ terms of the correct strength, as in Eq.~(\ref{eq:dual1}), (\ref{eq:dual2}).

Adiabatic computation is less suitable for qubits whose spectrum is nondegenerate since the thermal environment tends to relax the qubits to their individual ground state $\ket{0}$ instead of the joint multiple-qubit ground state of the adiabatic Hamiltonian. Thus other qubits with a degenerate spectrum such as flux-qubits currently used in quantum annealing could be considered for an implementation of adiabatic computation.

A direct realization using electrons and their spins may also be possible: at least one can imagine a spin-dependent hopping term realizing the Hadamard gate. In this direct realization the Hamiltonian requires a {\em strong attractive interaction} between neighboring electrons propagating over one-dimensional lines (in the adiabatic model there is in addition a repulsive interaction between next-nearest neighbor electrons across a plaquette, see Fig.~\ref{fig:connect}). One could consider using an particle/hole encoding alternating for adjacent wires so that particle and hole on nearest-neighbor wires attract via the Coulomb interaction (while repelling each other on next-nearest neighbor wires). Another idea is to use spin qubits localized in arrays of quantum dots and obtain the XX+YY coupling by letting two spin qubits
virtually exchange a boson via a quantum bus. Such a quantum bus can be a superconducting microwave resonator as in e.g. \cite{petersson:circuit} or
a standing surface acoustic wave as proposed in e.g. \cite{schuetz:transducer}. Strong ZZ coupling could again be obtained via an electrostatic interaction.

We emphasize that these proposed implementations are simply sketches of
how to attain the requisite interactions using existing technologies.   
Any quantum technology that allows the implementation of strong Ising
couplings and weak hopping terms allows the construction
of a Hamiltonian quantum computer in principle.    In the next
section we discuss various obstacles that stand in the way
of constructing such quantum computers in practice.



\section{Inaccuracy, Noise and Errors}


Any physical implementation of the construction given here will be susceptible to manufacturing
inaccuracies and to dynamic noise, which will induce errors in the
computation.  Instead of simulating a one-dimensional quantum circuit the Hamiltonian or adiabatic computation could simulate a fault-tolerant quantum circuit. In general such fault-tolerant quantum circuit, for example the surface code architecture \cite{quantumcodes_review}, is a 2D nearest-neighbor circuit. This implies that the time-independent Hamiltonian or adiabatic computation takes place on a 3D grid where the degrees of freedom are represented by a 2D-dimensional propagating membrane. Quantum measurements of subsets of qubits during the computation (to implement quantum error correction) can in principle be incorporated in any geometry by terminating the spatial regions of qubits which are to be measured so that they cannot propagate any further.  No rigorous mathematical analysis of adiabatic or Hamiltonian computation on such a 3D grid with regions in which partial membranes terminate early has been undertaken, although one expects similar (spectral) results to hold as in the 2D grid case. 

Running the encoded quantum circuit could provide protection against all errors or inacurracies in the Hamiltonian or adiabatic computation which can represented as errors on the qubits in the simulated fault-tolerant circuit as long as the induced errors have sufficient locality in space and time \cite{AGP:ft}. We discuss a few issues to be resolved. Consider what happens if the string dynamics is very slow so that a single-qubit noisy gate realized in some plaquette is repeatedly done and undone. It means that the error rate on this single-qubit gate will be high.  It was argued in \cite{GTV:adia} that the forward motion of the string is constant in the expanding region of the grid, but does this still hold for a noisy version of the string motion? 

Static disorder in the Feynman Hamiltonian leads to Anderson localization of the
clock variable \cite{Landauer90}.    While the extended nature of the
multi-clock string may render the Margolus Hamiltonian less susceptible
to localization, no analysis of localization in this system has been
performed.


Static or dynamic disorder in the ideal time-independent Hamiltonian or coupling to a finite temperature bath also leads to errors which represent {\em leakage errors} in the original encoded circuit. For example, in the dual rail encoding, amplitude damping on one of the qubits at a site leads to the state $\ket{00}$, i.e. the loss of a particle. Bitflip errors on the pair of qubits at a site can lead to the state $\ket{11}$ and thus again a leakage error. 

Another important source of errors (due to perturbative corrections, static disorder in the blocking terms or thermal activation of forbidden tunneling) leads to the string to become disconnected or incorrect. This can lead to the computation getting stuck: the forward motion of a particle $1$ is blocked as the particle, say, $2$, which gates this motion has already moved further in its computation. However the forward motion of particle $2$ will continue to be blocked while its backward motion, rejoining particle $1$, can lead to a lowering of its energy. Incorrect string states in which by some means a particle gets on the wrong CNOT track and picks up a $X$ can be surpressed by making a CNOT gate long. One can thus expect that a sufficiently low-density of string breaks or wrong string turns can be dealt with by simulating a fault-tolerant quantum circuit which includes leakage protection. Leakage errors are of course also an issue in circuit-based quantum computation requiring leakage reduction units such as quantum teleportation or swapping into fresh qubits (see e.g. \cite{AT:leakage} and references in \cite{quantumcodes_review}). 

\subsection{String Loops and Multiple-Time Wavefronts}
\label{sec:loopy}


Last but not least, we consider what happens when the number of particles on a line is not conserved, or in the dual-rail encoding, bit-flip errors generate new particles on the line. These are dangerous sources of errors which in the presented model can lead to lower energy states if we do not add additional penalty terms. If there is at most one particle per line the string always runs down the lattice as in Fig.~\ref{fig:rotated_grid}. However, if one allows for errors which create new particles, then one can create a string which loops back onto itself. The terms of $H_{\rm string}$ are such that when one particle is present on a site, it is favorable to have all nearest-neighbor sites be occupied with particles, so a string which comes from the top, loops back and closes onto itself will in fact have lower energy than the ground space of $H_{\rm string}$ in the single-particle per line sector (as on the looping string there is one site where a particle has 3 neighbor particles, elsewhere each particle has 2 neighbors). But we can add a term $H_{\rm no\; loop}=E \sum_{i,j} {\bf n}[i,j] {\bf n}[i+1,j+1]$ to the Hamiltonian where $E \gg \Delta$ which gives an energy penalty for a state which has two particles next to each other on a line. This term gives a direct energy penalty to a string looping back. The term does not prevent there from being another piece of string, i.e. a set of particles representing a new partial wavefront, that is created further away on the grid. However, an open string segment will pick up energy penalties at its boundaries due to $H_{\rm string}$ and a closed string segment will pick up energy with respect to $H_{\rm no\; loop}$. 

Energetically penalizing such errors is particularly relevant in the dual-rail encoding where the particles are represented by pairs of qubits and thus bitflip errors can create particles. Choosing $\ket{00}$ as the lowest-energy no-particle state is useful so qubit relaxation will not cause the creation of new particles. For qubits, the term $H_{\rm no\; loop}$ translates into a set of ZZ terms between the qubits horizontally across plaquettes.
It is an open question whether the physics that one obtains when such errors are suppressed but not eliminated, e.g. the splitting, creation and collisions of additional (closed or open) wave fronts, can be handled by error correction.

One expects that the membrane computation in 3D will be more robust than string computation in 2D. A piece of open membrane which has been disconnected from the rest of the membrane computation costs an energy scaling with the perimeter of the membrane, thus suppressing the formation of large open disconnected membranes. Closed membranes are made energetically unfavorable by using no-loop terms as in the string computational model presented in this paper.

\section{Conclusion}

This paper has shown how to perform universal Hamiltonian and adiabatic quantum
computation using only two-qubit interactions and without
higher-order perturbation gadgets.    Instead, 
energy penalties are used to enforce the desired 
Hamiltonian at lowest order. The resulting low-energy Hamiltonian could be used
either in a dynamic fashion or in an adiabatic/quantum annealing
context to encode the quantum computation in the ground state.
The construction given here was designed to be potentially
physically realizable using
large-scale quantum integrated circuits consisting of superconducting
qubits or quantum dots.    Many open questions remain, notably
how to deal with errors induced by manufacturing inaccuracy and by dynamic
noise.




\bigskip\noindent{\it Acknowledgements}

 BMT acknowledges funding through the European Union via QALGO FET-Proactive Project No. 600700.  BMT would like to think Ben Criger and Ida DiVincenzo for some help with the figures. This research was supported in part by Perimeter Institute for Theoretical Physics. Research at Perimeter Institute is supported by the Government of Canada through Industry Canada and by the Province of Ontario through the Ministry of Economic Development $\&$ Innovation. SL acknowledges
funding by Google, and would like to thank A. Bookatz, E. Farhi, L. Maccone
and the quantum information teams lead by H. Neven at
Google Research and by M. Amin at D-Wave for many helpful discussions.

\bibliography{refs_hamiltonian}

\appendix

\section{Rotating Away The Internal Dynamics}
\label{sec:rot}

We assume a Hamiltonian computation on the rotated grid with effective Hamiltonian $H^{\it eff}=g \sum_ p H_{\rm cond. hop}^p$ with $H_{\rm cond. hop}$ as in Eq.~(\ref{eq:cond_hop_CNOT}) inside CNOT regions and Eq.~(\ref{eq:effham1}) for single-qubit gates or $I$ wires, acting in the correct string subspace. Strings in this subspace can be parametrized as $\ket{z,{\bf s}}=\ket{z, s_1=\pm 1, \ldots s_{k(z)}=\pm 1}$ where $z$ is a $2m$-bit string representing the string degree of freedom and ${\bf s}$ is a set of $k(z)$ binary labels, one for each short CNOT region through which the string $z$ goes,  which indicate whether the correct string goes below or above the plane. The values $s_i=\pm 1$ thus correspond to the internal states of the control particles through which the string $z$ goes. Clearly, the number of these additional labels $k(z)$ depends on the string $z$, i.e. some strings do not have any additional labels.

We can write $P_{\rm correct\,string}=\sum_{z, {\bf s}}  \ket{z,{\bf s}}\bra{z,{\bf s}}$.

Let $V(z,{\bf s} | z',{\bf s'})$ be the transformation on the internal states which is executed by the circuit in going from a correct string 
$(z', {\bf s'})$ to another correct string $(z, {\bf s})$. In other words, $V$ is the total matrix-valued amplitude of the path from string $(z',{\bf s'})$ to $(z,{\bf s})$. We can write
\begin{equation}
H^{\it eff}=-g  \sum_{(z,{\bf s})} \sum_{(z',{\bf s'})\in {\rm Neigh}(z,{\bf s})} V(z,{\bf s} | z',{\bf s'})\ket{z,{\bf s}}\bra{z',{\bf s'}}.
\label{eq:adj}
\end{equation}
where ${\rm Neigh}(z,{\bf s})$ are the correct strings which are neighbors of $(z,{\bf s})$ meaning that conditional hopping terms in $H^{\it eff}$ can map $(z,{\bf s})$ onto $(z',{\bf s'})$ and vice versa. 

In addition, we have the composition $V(z,{\bf s}  | z'',{\bf s''})=\sum_{{\bf s'}} V(z,{\bf s} | z',{\bf s'}) V(z',{\bf s'}| z'',{\bf s''})$ where $V(z,{\bf s}  | z'',{\bf s''})$ is independent of the intermediate string $z'$: all paths from $(z'', {\bf s''})$ to $(z, {\bf s})$ execute the same computation independent through which intermediate string $z'$ they go. A transformation $V(z  | z'',{\bf s''})$ is unitary on the internal state space if the string $z$ has no control particle labels, hence we do not post-select the dynamics on the final state of some control particles. It also holds that $V^{\dagger}(z,{\bf s}  | z',{\bf s'})=V(z',{\bf s'}  | z,{\bf s})$.

We write down an isometry $W$ such that $W^{\dagger} H^{\it eff} W$ is a transition matrix in the string subspace $\ket{z}$ independent of the internal dynamics and the spin-labels ${\bf s}$: 
\begin{equation}
W=\sum_{z,{\bf s}} V(z,{\bf s} | z_{\rm init})    \ket{z,{\bf s}}\bra{z},
\end{equation}
so that $W^{\dagger} W=\sum_z \ket{z}\bra{z}=P_{\rm string}$. Note that $W$ is an isometry as it maps from the string space to a higher-dimensional correct string space where the string has additional labels. Using that $\sum_{\bf s}V(z,{\bf s}| z',{\bf s'}) V(z',{\bf s'}| z_{\rm init})=V(z,{\bf s} | z_{\rm init})$, $V(z|z)=I$ etc., one obtains 
\begin{equation}
W^{\dagger} H^{\it eff} W=-g \sum_{z, z'\in {\rm Neigh}(z)} \ket{z}\bra{z'},
\end{equation}
representing only the string dynamics on the grid. 

In the quantum adiabatic computation the effective Hamiltonian, Eq.~(\ref{eq:circuitH})  is slightly different than the Hamiltonian in Eq.~(\ref{eq:adj}). It includes a diagonal term of the general form $\sum_{z,{\bf s}} \alpha_z \ket{z,{\bf s}}\bra{z,{\bf s}}$ where the weights $\alpha_z$ do not depend on the internal states of the control particles ${\bf s}$. Conjugation with $W$ of such term gives $W^{\dagger} \sum_{z,{\bf s}} \alpha_z \ket{z,{\bf s}}\bra{z,{\bf s}} W=\sum_{z} \alpha_z \ket{z}\bra{z}$, again removing the internal control particle labels. In addition, the effective Hamiltonian includes a diagonal term $H_{\rm input}$ which penalizes the internal states of particles before they run through the computation: this means that the strings $(z,{\bf s})$ for which this term is non-zero are always strings $z$ without control-particle labels and $V(z,{\bf s}|z_{\rm init})=V(z|z_{\rm init})=I$, or the dynamics from $z_{\rm init}$ to these strings $z$ is trivial. This implies that conjugation by $W$ leads to a term which penalizes the internal state of particles before the computation is executed, identical to the effect of such term in \cite{GTV:adia}.

In all, what these arguments show is that even for a circuit with CNOT and Toffoli gates, one can transform away the unitary dynamics of the computation and obtain a model which purely represents the string dynamics. One can then invoke the previous results which analyze the string dynamics in the case of Hamiltonian computation (XY model) or quantum adiabatic computation (ferromagnetic XXZ chain with kink boundary conditions, Eq.~(\ref{eq:XXZ})). 

It can be noted that the inclusion of higher-order terms in the perturbative expansion (of $O(g^{k+1}/\Delta^{k})$) in the correct string subspace do not affect the validity of this analysis. Such higher-order terms, due to multiple hopping, lead to new terms in Eq.~(\ref{eq:adj}) of the form $\gamma_{z,z'} V(z,{\bf s} | z',{\bf s'})\ket{z,{\bf s}}\bra{z',{\bf s'}}$ for some coefficients $\gamma_{zz'}$ where $(z,{\bf s})$ and $(z',{\bf s'})$ are not direct neighbors, but are connected through multiple hops.


\end{document}